# Itinerant ferromagnetism mediated by giant spin polarization of metallic ligand band in van der Waals magnet $Fe_5GeTe_2$

K. Yamagami[1,2], Y. Fujisawa[1], B. Driesen[1], C. H. Hsu[1,3], K. Kawaguchi[2], H. Tanaka[2], T. Kondo[2], Y. Zhang[4],

H. Wadati[5], K. Araki[6], T. Takeda[6], Y. Takeda[7], T. Muro[8], F. C. Chuang[3,9,10],

Y. Niimi[11], K. Kuroda[2], M. Kobayashi[6,12], and Y. Okada[1]

[1]Okinawa Institute of Science and Technology Graduate University,

Tancha, Onna-son, Kunigami-gun Okinawa, 904-0495, Japan.

[2]Institute for Solid State Physics, The University of Tokyo, Kashiwa 277-8581, Japan.

[3]Department of Physics, National Sun Yat-sen University, Kaohsiung 80424, Taiwan

[4]Institute of High Energy Physics, Chinese Academy of Sciences, Shijingshan District, Beijing, 100049, China.

[5]Graduate School of Material Science, University of Hyogo, Ako, Hyogo 678-1297, Japan.

[6]Department of Electrical Engineering and Information Systems, Graduate School of Engineering,

The University of Tokyo

[7]Materials Sciences Research Center, Japan Atomic Energy Agency, 1-1-1 Kouto, Sayo-cho, Sayo-gun,

Hyogo 679-5148, Japan

[8]Japan Synchrotron Radiation Research Institute (JASRI), 1-1-1 Kouto, Sayo, Hyogo 679-5198, Japan

[9]Department of Physics, National Tsing Hua University, Hsinchu 30013 Taiwan

[10]Physics Division, National Center for Theoretical Sciences, Hsinchu 30013, Taiwan

[11]Department of Physics, Graduate School of Science, Osaka University, Japan.
[12]Center for Spintronics Research Network, The University of Tokyo, Hongo, Bunkyo-ku, Tokyo 113-8656, Japan.

**ABSTRACT**

We investigate near-Fermi-energy ($E_F$) element-specific electronic and spin states of ferromagnetic van der Waals (vdW) metal $Fe_5GeTe_2$. The soft x-ray angle-resolved photoemission spectroscopy (SX-ARPES) measurement provides spectroscopic evidence of localized Fe $3d$ band. We also find prominent hybridization between the localized Fe $3d$ band and the delocalized Ge/Te $p$ bands. This picture is strongly supported from direct observation of the remarkable spin polarization of the ligand $p$ bands near $E_F$, using x-ray magnetic circular dichroism (XMCD) measurements. The strength of XMCD signal from ligand element Te shows the highest value, as far as we recognize, among literature reporting finite XMCD signal for none-magnetic element in any systems. Combining SX-ARPES and elemental selective XMCD measurements, we collectively point an important role of giant spin polarization of the delocalized ligand Te states for realizing itinerant long-range ferromagnetism in $Fe_5GeTe_2$. Our finding provides a fundamental elemental selective view-point for understanding mechanism of itinerant ferromagnetism in low dimensional compounds, which also leads insight for designing exotic magnetic states by interfacial band engineering in heterostructures.

Understanding the mechanism of ferromagnetism in metals has been a long-standing non-trivial issue in wide variety of condensed matter systems. Based on conventional itinerant picture, the so-called Stoner model explains mechanism of itinerant ferromagnetism [1]. For example, in the transition metal Ni, a spin polarized band with exchange band splitting emerges below the Currie temperature ($T_C$) [2,3]. This picture supports ferromagnetism arisen from simple itinerant Ni $3d$ state. On the other hand, Fe shows persistent exchange band splitting above $T_C$ [4]. This picture supports the localized nature of Fe $3d$ states rather than simple itinerant picture. In complexed ferromagnetic metal compounds, the situation becomes richer [5,6]. For example, MnP shows band deformation across $T_C$, in line with Stoner picture, with non-trivial pseudo-gap formation near Fermi energy ($E_F$) at low temperature [5]. In $SrRuO_3$, a violation of Stoner picture is reported, with band renormalization with strong coupling to bosonic mode near $E_F$ [6]. Regardless of material dependent complicates and richness, the most essential question for understanding ferromagnetism in metals has been clarifying whether orbital with magnetic moment should be viewed as localized or itinerant picture in many cases.

Owing to rapidly increasing interest in searching for novel magnetism in van der Waals (vdW) compounds, one particularly important challenge is understanding mechanism of itinerant ferromagnetism with reduced dimension [7,8,9,10,11,12]. Recent studies in vdW ferromagnetic metal $Fe_3GeTe_2$ ($T_C$ ~220 K) supports localized nature of Fe $3d$ states [13]. However, this picture is not easily understood based on so called Mermin–Wagner theorem since this theorem does not allow long-range ferromagnetism starting from localized isotropic Heisenberg spins in two-dimensions [14]. To reconcile, investigation on vdW ferromagnetic insulators $Cr_2Ge_2Te_6$ [15,16,17,18,19] and $CrI_3$ [20,21,22] provide useful insight. In these systems, the ligand states significantly contribute to the ferromagnetism through hybridization with orbital having magnetic moment [18,19,21,22]. Thus, this information provides a motivation of clarifying metallic vdW ferromagnet with an elemental selective point of view, particularly focusing on a role of ligand state. However, lacking direct investigation of elemental specific electronic and spin states hinders understanding of the mechanisms of emergent ferromagnetism in metallic vdW ferromagnets so far.

Here, we investigate the elemental selective valence-band electronic and spin states of $Fe_5GeTe_2$, using soft x-ray angle-resolved photoemission spectroscopy (SX-ARPES) [23] and soft x-ray magnetic circular dichroism (XMCD) in x-ray absorption spectroscopy (XAS) [24]. $Fe_5GeTe_2$, we focus on, is especially important itinerant ferromagnetic vdW compound due to the its highest $T_C$ ~310 K within existing metallic vdW ferromagnet [25,26,27]. The crystal structure of $Fe_5GeTe_2$ (as in Fig. 1(a)) is a hexagonal lattice with three $Fe_5GeTe_2$ layers in unit cell. This report provides the first clarification of elemental specific near $E_F$ electronic and spin states in $Fe_5GeTe_2$. Our findings coherently point that the remarkably large spin polarized metallic ligand Ge/Te states play a major role for mediating itinerant long-range ferromagnetism in this vdW compound, providing a fundamental view-point for understanding of the emergence of itinerant ferromagnetism in low dimensional systems.

A quenched-single crystal $Fe_5GeTe_2$ was synthesized in an evacuated quartz tube with an $I_2$ transport agent, as described in literature [26]. The temperature-dependent magnetic susceptibility measurement gives $T_C \approx 310$ K for our sample. The SX-ARPES measurements were performed at BL25SU of SPring-8 [28], and the incident photon energy ($h\nu$) was set to 450 eV with energy resolution of ~50 meV. Thanks to the smaller beam spot size, the observed band dispersion in SX-ARPES is slightly sharper than VUV-ARPES by He discharge lamp, related to the $\sqrt{3} \times \sqrt{3}$ domain structure [26], even if the energy resolution is much better for the latter case (see Fig. S1 in the supplemental materials). The XAS and XMCD measurements were performed at BL23SU of SPring-8 [29]. The details of experimental conditions for

SX-ARPES, XAS, and XMCD are described in the supplemental materials.

We first present capability of orbital selective photoemission spectroscopies (PES), based on $h\nu$ dependence of the photoionization cross-section ($\sigma_{nl}$) and density functional theory (DFT) based partial density of states (PDOS) calculations. Fig. 1(b) shows PES spectrum with both $h\nu$ = 1486.6 eV (black line with empty symbol) and $h\nu$ = 450 eV (purple line filled symbol). For both spectra, we observe three characteristic peak features labelled as A ($E$-$E_F \approx$ -0.5 eV), B ($E$-$E_F \approx$ -2.8 eV), and C ($E$-$E_F \approx$ -4.3 eV). However, relative peak intensity largely depends on incident photon energy $h\nu$. In PES with $h\nu$ = 1486.6 eV, peak feature C dominates rather than A and B, and overall spectral shape is similar to calculated PDOS for Ge $4p$ and/or Te $5p$ [see blue and green curves in Fig. 1(c)]. On the other hand, spectrum with $h\nu$ = 450 eV shows predominant peaks A and B, and overall spectral shape matches very well with calculated PDOS for Fe $3d$ [see red curve in Fig. 1(c)]. These orbital selective PES spectra can be consistently understood based on $h\nu$ dependence of $\sigma_{nl}$. At $h\nu$ = 1486.6 eV, $\sigma_{nl}$ for the Fe $3d$, Te $5p$ and Ge $4p$ are $\sigma_{Fe3d}/\sigma_{Te5p} \simeq 0.55$ and $\sigma_{Fe3d}/\sigma_{Ge4p} \simeq 0.97$, while, at $h\nu$ = 450 eV, their relation represents $\sigma_{Fe3d}/\sigma_{Te5p} \simeq 4.7$ and $\sigma_{Fe3d}/\sigma_{Ge4p} \simeq 10$ [30]. Therefore, the Ge/Te derived signal reasonably becomes more prominent using $h\nu$ = 1486.6 eV whereas Fe $3d$-derived signature becomes more prominent with $h\nu$ = 450 eV.

The experimental SX-ARPES ($hv$ = 450 eV) at 50 K simply captures the most essential electronic feature seen in DFT calculation. Fig. 1(e) shows calculated relative spectral weight contribution between $d$ and $p$ states on high symmetry line along $\bar{\Gamma}$ - $\bar{M}$ direction [the yellow arrow in Fig. 1(d)]. Here, as in Fig. 1(c), $d$ and $p$ contribution mainly originate from Fe 3$d$ and Te 5$p$ derived states, respectively. The photoemission intensity image along the $\bar{\Gamma}$-$\bar{M}$ direction is shown in Fig. 1(f). Less dispersive spectral weight distribution is seen (plotted with circle) around $\bar{\Gamma}$ and $\bar{M}$ points, with disconnected spectral weight distribution around characteristic momentum $|k_H|$ = 0.38 Å$^{-1}$. On the other hand, metallic hole-like band is also observed around $\bar{\Gamma}$ point (plotted with triangle). Combining SX-ARPES and DFT calculation, experimental flat distribution of spectral weight can be assigned as convolution of *multiple* Fe 3$d$ bands. Also, hole band around $\bar{\Gamma}$ point is assigned as Ge/Te $p$ band. More pronounced intensity for Fe 3$d$ bands than delocalized Ge/Te $p$ hole bands is consistently understood due to cross-section $\sigma_{nl}$ at $hv$ = 450 eV. Furthermore, as shown by vertical broken lines in Fig. 1(e), experimental observation of disrupted spectral weight distribution across ±$k_H$ is qualitatively consistent with prominent hybridization around this momentum [31].

The elemental selective XMCD measurement is particularly powerful for magnetic systems with complicated band structure, such in Fe$_5$GeTe$_2$. The XMCD signal is proportional to moment $M$ for magnetic element. For none-magnetic element, XMCD signal is proportional to the number of electrons with opposite spins $N(\uparrow_{major})$- $N(\downarrow_{minor})$ from spin-polarized Ge/Te derived bands. If the Ge/Te $p$ band magnetically couples with the flat Fe 3$d$ band through the hybridization, the itinerant Ge/Te $p$ derived orbitals near $E_F$ are expected to provide finite XMCD signals.

Based on selection rule and orbital nature, the Fe $L_{2,3}$-edge corresponds to a 2$p$ → 3$d$ excitation process. Then, the magnetic signal from Fe 3$d$ derived states is first presented from Fe $L_{2,3}$-edge XAS and XMCD spectra, whose experimental geometry are shown in Fig. 2(a). Figure 2(b) shows the Fe $L_{2,3}$-edge XAS ($\mu^+$ and $\mu^-$) and XMCD ($\mu^-$–$\mu^+$) spectra at 20 K under a magnetic-field of 6 T. The $\mu^+$($\mu^-$) denotes the XAS absorption intensity for parallel (anti-parallel) alignment of the photon helicity and the sample magnetization direction. The metallic XAS spectral shape is consistent with hybridization driven finite Fe 3$d$ derived states at $E_F$ [29,32], which is further supported by core level PES spectra (see Fig. S2 in the supplemental materials). Figure 2(c) shows the magnetic field dependence of the XMCD intensity at the photon energy of 707.1 eV. Note here that three different Fe-site exists in Fe$_5$GeTe$_2$, as shown in Fig. 1(a). We find that the XMCD signal was overlapped from all three Fe sites in energy, and the signal is expected to be integral from three Fe sites. Indeed, the elemental-specific magnetization curve shape shown in Fig. 2(c) represents small coercive field with saturation field 0.6 T, which is similar to magnetic curve seen in bulk measurement [26,27]. In addition to the calculated orbital and spin moment of $m_{orb}$ = 0.1 $\mu_B$/Fe and $m_{spin}$ = 1.8 $\mu_B$/Fe, the estimated total moment of $m_{total}$ = $m_{orb}$ + $m_{spin}$ = 1.9 $\mu_B$/Fe for 6 T, using the XMCD sum-rules [32,33,34] (see Fig. S3 in supplemental materials), is nearly identical to that from macroscopic magnetometry [27].

Spin-polarized signal from Te/Ge derived orbitals have also been observed by XMCD. Figures 3(a) and (b) represent the Te $M_{4,5}$-edge Ge $L_{2,3}$-edge XAS and XMCD spectra taken at 20 K under a magnetic field of 6 T. Based on selection rule and orbital natures, the Te $M_{4,5}$-edge and Ge $L_{2,3}$-edge can be regarded as excitations of 3$d$ → 5$p$ and 2$p$ → 4$s$/4$d$, respectively. The magnetic field dependence of the XMCD signal for Te (at 572.4 eV labeled as X) and Ge (~1215.1 eV and ~1219.3 eV, labeled as Y and Z) are shown in Figs. 3(c) and (d), respectively. This is the first observation of spin-polarized ligand Ge/Te states in Fe-Ge-Te vdW ferromagnets. Importantly, one can further see the sign change of the XMCD signal depending on elements and their excitation process.

To confirm the hybridization picture, we plot the magnetic field dependence of XMCD signals normalized at the saturated value of each absorption edges in Fig 4(a). Note that the magnetic behavior of Ge/Te states are qualitatively identical to that of the Fe state except for the sign difference, indicating the strong magnetic coupling between the Fe and Ge/Te states. The XMCD signal from Fe $L_{2,3}$-edge dominantly probes 3$d$ states because of the dipole transition with different azimuthal $\Delta l = l_{final} - l_{initial}$ = +1. If the spin alignment of different orbitals in two elements is anti-parallel, the sign of XMCD signals should be opposite, as long as the sign of $\Delta l$ for the two excitations is the same. The XMCD signal from the Te $M_{4,5}$-edge mainly detects excited 5$p$ states with $\Delta l$ = -1 and its sign is the same as that of the Fe $L_{2,3}$-edge. This means that the spin polarization of the Te 5$p$ state is anti-parallel to the Fe 3$d$ states. On the other hand, the XMCD signal for the Ge $L_{2,3}$-edge has two signals with different signs (assigned as Y and Z). This can be coherently understood by considering different $\Delta l$ involved in the two-excitation process, since the sign of the XMCD signals can be flipped with opposite sign of $\Delta l$. Compared with the previous XMCD studies for the Ge $L_{2,3}$-edge [35], the XMCD signals at Y and Z are from the final states of 4$s$ ($\Delta l$ = -1) and 4$d$ ($\Delta l$ = +1), respectively. Therefore, in addition to Te-5$p$, Ge-4$s$/4$d$ also has spin-polarized states with its spin alignment anti-parallel to those of Fe 3$d$ under external magnetic field, which is summarized schematically in Fig. 4(b). While macroscopic moment originates mainly from localized Fe spins, such XMCD signal points that spin-polarized metallic ligand band should play a major role for mediating itinerant ferromagnetism.

Surprisingly large XMCD signal from Te $M_5$-edge is an important fact. The strength of XMCD signal is defined as $(\mu^- - \mu^+)/(\mu^- + \mu^+ - B.G.)$, and the extracted value for Fe, Ge, Te signal from ligand Te $5p$ reaches to the value of Fe $L_3$-edge. So far, the similarly defined finite XMCD signal strength from non-magnetic element has been reported in various literatures [22,36,37,38,39,40,41,42,43,44,45]. However, as far as we recognize, our value of ~30% XMCD signal for non-magnetic element is much larger than any of previous reports for different materials unlimited in vdW coupled systems. Our findings collectively point that the spin polarization of the delocalized ligand states plays a major role for realizing itinerant long-range ferromagnetism in low dimensional systems. Indeed, validity of this picture is strongly supported by direct observation of band hybridization and the induced remarkable spin polarization for itinerant ligand Ge/Te states [Fig. 4(a)]. Since spin-polarized density of states for ligand states $N(\uparrow_{major}) - N(\downarrow_{minor})$ at $E_F$ is expected to be higher when the hybridized flat band is located near $E_F$, our observation of the flat band energy at $E-E_F \sim -0.3$ eV [see Fig. 1(f)], which is closer to $E_F$ than that of $Fe_3GeTe_2$ ($E-E_F \sim -0.4$ eV [13]), can explain the higher $T_C$ of $Fe_5GeTe_2$ than that of $Fe_3GeTe_2$.

Interesting consequence of observed large XMCD signal for heavy element Te is possible major role of spin-orbit (SO) coupling effect for determining detailed arrangement of Fe derived spins. This hypothesis is in line with the conclusion drawn, for example, in itinerant ferromagnetic system intermetallic iron-based compounds FePt. In this system, large spin polarization of non-magnetic heavy element Pt induced by Fe introduces prominent SO coupling effect to determine spin structure of Fe including magnetic anisotropy [46,47,48]. As also in vdW ferromagnetic insulators $Cr_2Ge_2Te_6$ [15,16,17,18,19] and $CrI_3$ [20,21,22], we think SO coupling plays a major role to determine detailed arrangement of localized spins of Fe in present vdW ferromagnetic metal $Fe_5GeTe_2$. The information obtained in this study will be essential for understanding of temperature dependent unique physical properties, instead of simple ferromagnetism, including formation of ferrimagnetism, Fano-shaped near $E_F$ electronic state, none collinear spin structure, and their exotic temperature evolution of domain structuring [26,27,49,50]. In particular, we think Fe(1) plays an important role for controlling physical properties in $Fe_5GeTe_2$ through hybridization with heavy ligand state Te since this magnetic sites is the nearest neighbor for Te site [see Fig. 1(a)] and missing in related material $Fe_3GeTe_2$ (see Fig. S4 in the supplemental materials). The direct visualization of the electronic structure at the Fe-Te hybridized layer would bring a further insight on the vdW ferromagnets in $Fe_5GeTe_2$, that will be an interesting future study.

In summary, combining ARPES, DFT, and direct observation of elemental selective XMCD measurements, we provide collective picture of emergent hybridization driven spin polarization of none-magnetic ligand states. Our discovery of surprisingly large spin polarization of the delocalized ligand Te states points a major role for realizing itinerant long-range ferromagnetism in low dimensional systems, while difficulty of aligning localized Fe $3d$ spins by themselves in low dimensional system is expected. Beyond conventional magnetism, controlling magnetism through heavy ligand element is interesting root for designing emergent itinerant ferro magnetism in low dimensional systems, including their heterostructures.

**ACKNOWLEDGMENT**

We thank M. Kitamura for supporting the XAS/XMCD measurements. The SX-ARPES measurement was performed under the approval of BL25SU at SPring-8 (Proposals No.2019B1097). The XAS/XMCD measurements were performed under the approval of BL23SU at SPring-8 (Proposals No. 2019B-E16, A-19-AE-0036, 2019B3845, 2019B-E20, A-19-AE-0040, and 2019B3841). This work is partly supported by Japan Society and Science and Technology Agency (JST) Core Research for Evolution Science and Technology (CREST), Japan, Grant number JPMJCR1812. This work was partially supported by JSPS KAKENHI (Grant No.19H02683, 19H01816, and 19H05824), Japan. This work was partially supported the Spintronics Research Network of Japan (Spin-RNJ). Y. Z. acknowledges the Basic Research Funding of IHEP, CAS (Grant No. Y951556). F. C. C. and C. H. H. acknowledge support from the National Center for Theoretical Sciences and the Ministry of Science and Technology of Taiwan under Grants No. MOST-107-2628-M-110-001-MY3, 108-2911-I-110-502, and 107-2911-I-110-506. They also grateful to the National Center for High-performance Computing for computer time and facilities.

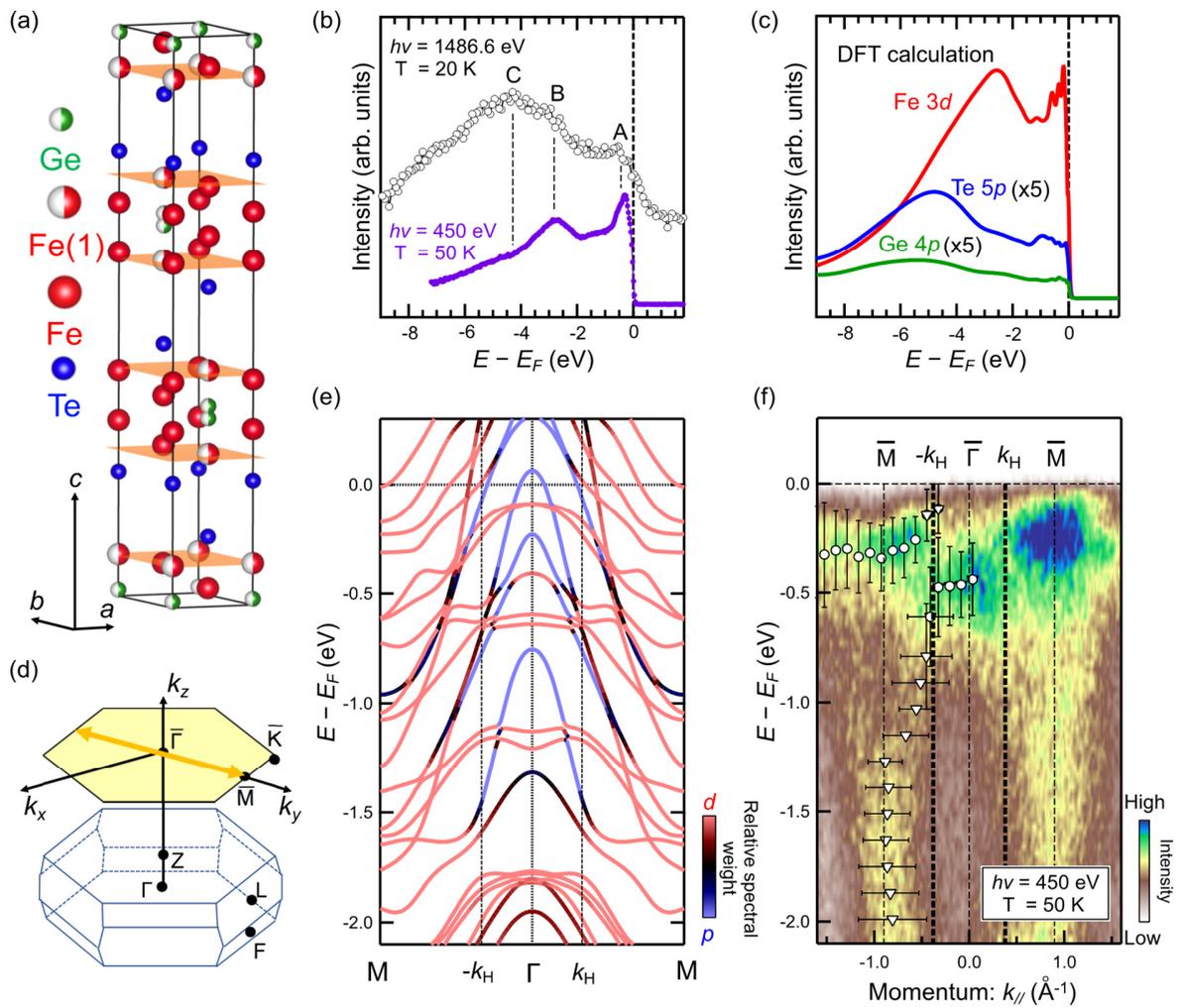

**Figure 1.** (Color online) (a) Crystal structure of $Fe_5GeTe_2$. It is know that 50% occupation for Fe(1) and Ge sites [27]. (b) The valence band PES spectra at the incident photon energy ($h\nu$) of 1486.6 eV and 450 eV. See main body for details of the spectral features indicated by A-C. (c) The DFT-calculated partial density of states, multiplied by Fermi Dirac function. (d) The bulk and projected Brillouin zones of $Fe_5GeTe_2$. (e) The DFT calculation for relative spectral weight between $d$ and $p$ orbitals. (f) Photoemission intensity plot along $\bar{\Gamma}$-$\bar{M}$ at 50 K and $h\nu$ = 450 eV.

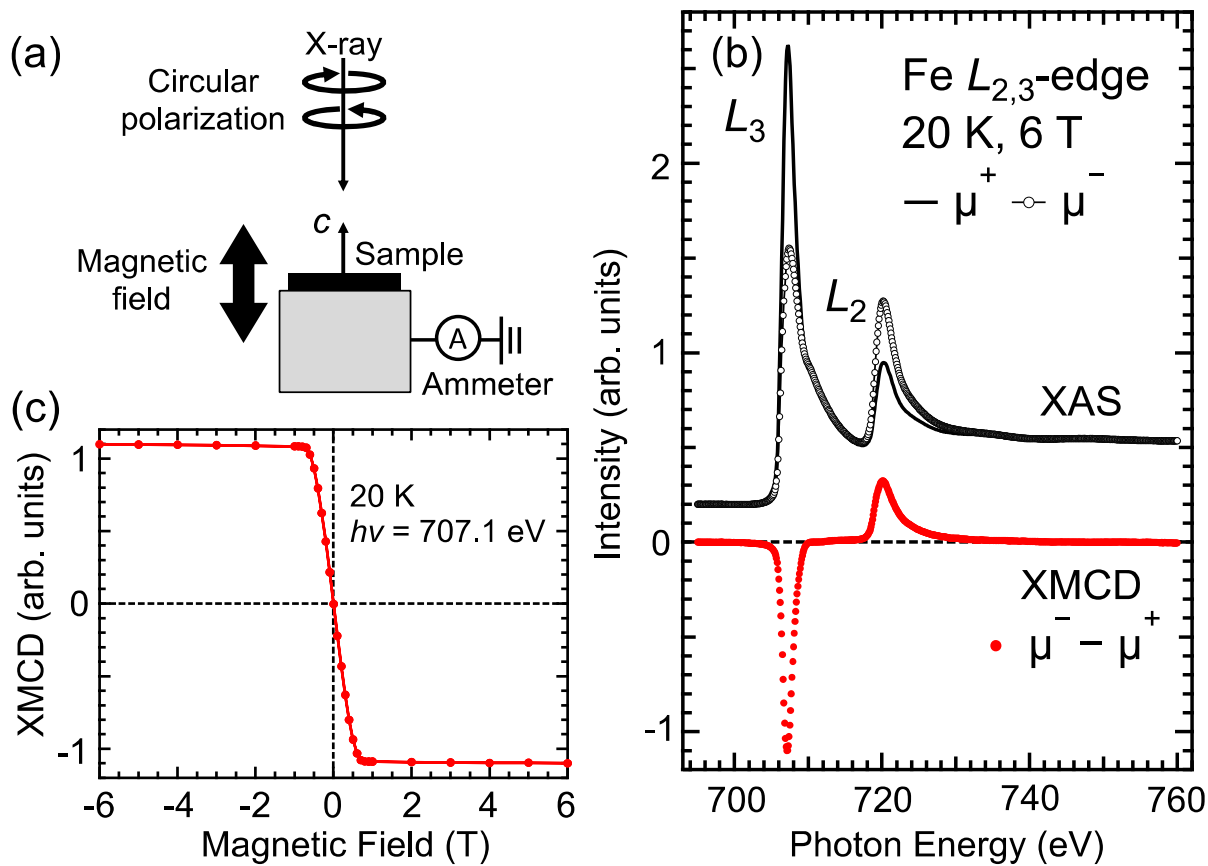

**Figure 2.** (Color online) (a) The experimental geometry of XAS/XMCD measurements in TEY mode. See method part for details. (b) The Fe $L_{2,3}$-edge XAS and XMCD spectra at 20 K under 6 T. The $\mu^+$($\mu^-$) denotes the XAS absorption intensity for parallel (anti-parallel) alignment of the photon helicity and the sample magnetization direction. (c) The magnetic-field dependence of the XMCD intensity of Fe $L_3$ (707.1 eV) at the photon energy of the XMCD peak.

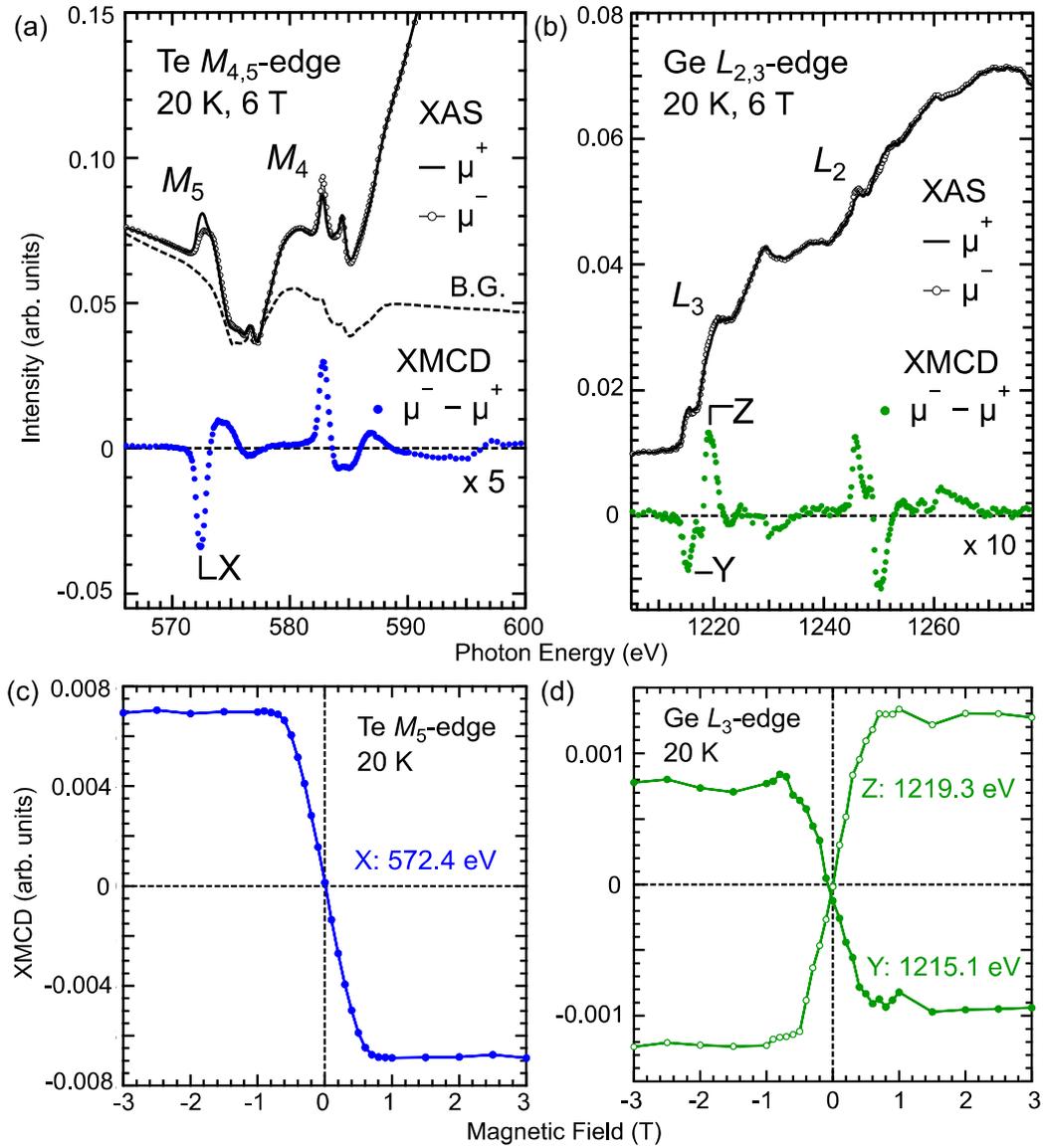

**Figure 3.** (Color online) (a,b) The Te $M_{4,5}$-edge and Ge $L_{2,3}$-edge XAS and XMCD spectra of Fe$_5$GeTe$_2$ at 20 K under 6 T. In Te $M_{4,5}$-edge, the background spectrum including in the Cr $L_{2,3}$-edge (≈576 eV and 585 eV) absorption from the focusing mirrors are shown as the black solid line. (c,d) The element-specific magnetization curves of Te and Ge with the photon energy set at the XMCD peaks; X (572.4 eV), Y (1215.1 eV), and Z (1219.3 eV).

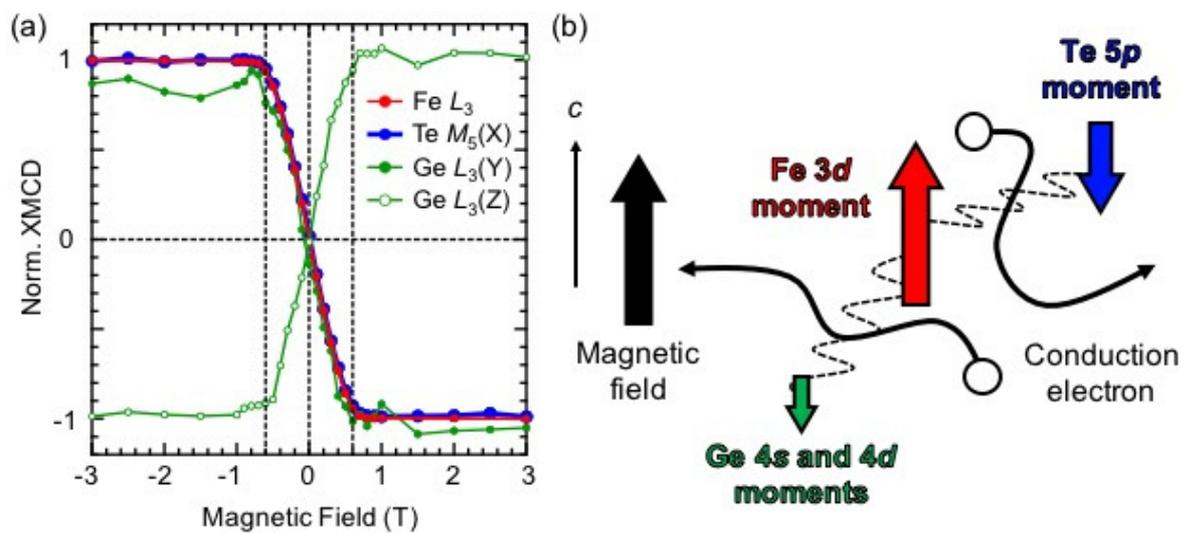

**Figure 4.** (Color online) (a) The magnetic-dependence of XMCD signals normalized at the saturated value of each absorption edges. (b) The sketch of the relationship about magnetic moments between Fe $3d$ and seemingly non-magnetic elements revealed by SX-ARPES and XMCD measurements.